\documentclass{emulateapj}
\usepackage{natbib}
\usepackage{graphicx}
\usepackage{amsmath}
\usepackage{float}
\newcommand{\sax} {SAX J2103.5$+$4545}
\newcommand{\saxs} {J2103}
\newcommand{\nh} {$N_{\text{H}}$}
\newcommand{\chisq} {$\chi^{2}$}
\newcommand{\delchisq} {$\Delta \chi^{2}$}
\newcommand{\nustar} {\textit{NuSTAR}}

\shorttitle{A phase dependent absorption feature in SAX J2103.5}
\shortauthors{Brumback et al.}

\slugcomment{Accepted for publication in The Astrophysical Journal}

\begin{document}

\title{A possible phase dependent absorption feature in the transient X-ray pulsar SAX J2103.5$+$4545}

\author{Brumback, M.C.{\altaffilmark{1}}, Hickox, R.C.{\altaffilmark{1}}, F\"urst, F.S.{\altaffilmark{2}}, Pottschmidt, K.{\altaffilmark{3,4}}, Hemphill, P.{\altaffilmark{5}}, Tomsick, J.A.{\altaffilmark{6}}, Wilms, J.{\altaffilmark{7}}, Ballhausen, R.{\altaffilmark{7}} }

\altaffiltext{1} {Department of Physics \& Astronomy, Dartmouth College, 6127 Wilder Laboratory, Hanover, NH 03755, USA} 
\altaffiltext{2} {European Space Astronomy Centre (ESA/ESAC), Operations Department, Villanueva de la Ca$\tilde{\text{n}}$ada Madrid, Spain} 
\altaffiltext{3} {CRESST, Department of Physics and Center for Space Science and Technology, UMBC, Baltimore, MD 210250, USA} 
\altaffiltext{4} {NASA Goddard Space Flight Center, Code 661, Greenbelt, MD 20771, USA} 
\altaffiltext{5} {Massachusetts Institute of Technology, 77 Massachusetts Ave, Cambridge, MA 02139, USA} 
\altaffiltext{6} {Space Sciences Laboratory, University of California, Berkeley, 7 Gauss Way, Berkeley, CA 94720, USA} 
\altaffiltext{7} {Dr. Karl Remeis-Sternwarte and Erlangen Centre for Astroparticle Physics, Sternwartstrasse 7, 96049 Bamberg, Germany}

\begin{abstract}
We present an X-ray spectral and timing analysis of two \nustar\ observations of the transient Be X-ray binary \sax\ during its April 2016 outburst, which was characterized by the highest flux since \textit{NuSTAR}'s launch. These observations provide detailed hard X-ray spectra of this source during its bright precursor flare and subsequent fainter regular outburst for the first time. In this work, we model the phase-averaged spectra for these observations with a negative and positive power law with an exponential cut-off (NPEX) model and compare the pulse profiles at different flux states. We found that the broad-band pulse profile changes from a three peaked pulse in the first observation to a two peaked pulse in the second observation, and that each of the pulse peaks has some energy dependence. We also perform pulse-phase spectroscopy and fit phase-resolved spectra with NPEX to evaluate how spectral parameters change with pulse phase. We find that while the continuum parameters are mostly constant with pulse phase, a weak absorption feature at $\sim\!12$ keV that might, with further study, be classified as a cyclotron line, does show strong pulse phase dependence.
\end{abstract}

\section{Introduction}

Accretion onto magnetized neutron stars provides a unique laboratory to study the behavior of gas in high magnetic and gravitational fields. In a binary consisting of a neutron star and a stellar companion, gas typically leaves the companion star through either Roche lobe overflow or a stellar wind (\citealt{reig2011}). If the gas has a low enough angular momentum, the neutron star can gravitationally trap part of this matter into an accretion disk. However, at the magnetosphere of the neutron star, the magnetic pressure exceeds the ram pressure of the disk and accretion is confined to the magnetic field of the neutron star (e.g.\ \citealt{wolff2016}). Such magnetically dominated accretion is not well understood, and yet it plays an important role in accretion onto magnetized objects such as neutron stars, white dwarfs, and young stars. Accretion powered pulsars offer unique opportunities to observe magnetically dominated accretion.

SAX J2103.5$+$4545 (hereafter J2103) is a Be X-ray binary (BeXB) consisting of a neutron star and a high mass stellar companion. It was discovered in 1997 by the \textit{Beppo}SAX satellite and observed to have a spin period of 358.61 $\pm$ 0.03 s (\citealt{hullemann1998}). The optical counterpart is a B IVe-Ve star with a visual magnitude of 14.2 (\citealt{reig2004}). Optical data suggest the distance to \saxs\ is $\sim\!6.5$\,kpc (\citealt{filippova2004,reig2004}), while X-ray data yield a smaller value of $\sim\!4.5$\,kpc (\citealt{baykal2007}). \saxs\ has an orbital period of 12.68 days and an eccentricity of 0.4, making it the BeXB with the shortest known orbital period (\citealt{baykal2000,baykal2007,reig2010}).

After its discovery in 1997, \saxs\ was detected in outburst by RXTE in 1999 (\citealt{baykal2000}). The 2--25 keV spectrum was described by an absorbed power law with a photon index of 1.27 $\pm$ 0.14 and \nh\ of (3.80 $\pm$ 0.10) $\times$ 10$^{22}$ cm$^{-2}$ (\citealt{baykal2002}). Additionally, \saxs\ was found to be in a spin up state with a rate of (2.50 $\pm$ 0.15) $\times$ 10$^{-13}$ Hz s$^{-1}$ (\citealt{baykal2000}). \cite{baykal2002} concluded that  discovery of this period derivative in conjunction with X-ray activity implied the formation of an accretion disk around the compact object.

X-ray outbursts have been detected by \textit{INTEGRAL} every 2--3 years since 2002 (\citealt{lutovinov2003,sidoli2005,ducci2008,reig2010,sguera2012,ducci2014}). During these events, a strong precursor flare with a flux of $\sim$\! 100 mCrab is followed by $\sim$100 days of regular outbursts where the flux is typically 20--40 mCrab. Regular outbursts occur on a 12 day cycle, consistent with periastron passage, and the peak X-ray intensity is an order of magnitude above that of quiescence (\citealt{reig2010}). These outbursts coincide with increased optical activity from the Be companion, including the H$\alpha$ line seen in emission (\citealt{reig2004,manousakis2007}). 

The existence of an H$\alpha$ emission line, in addition to optical and IR excess, indicates that a disk of ejected atmospheric material has formed around the Be companion star (\citealt{camero2014,reig2014}). Although the formation mechanisms of such a circumstellar disk are still uncertain, Keplerian disks supported by viscosity have been found to exist in isolated Be stars (\citealt{rivinius2013}). It is generally thought that \saxs\ spends 2--3 years in quiescence due to the formation time scale for this circumstellar disk (\citealt{camero2014}). When the disk reaches a critical size, the neutron star rapidly accretes much of the disk, resulting in the precursor flare. The remainder of the disk is accreted during subsequent periastron passages of the neutron star (\citealt{camero2014}).

While previous studies have examined the soft X-ray spectral properties of \saxs\ in outburst (e.g.\ \citealt{camero2014}), \cite{baykal2002} noted that spectrum of \saxs\ above approximately 10 keV becomes harder when the X-ray flux is higher. The launch of the Nuclear Spectroscopic Telescope Array (\textit{NuSTAR}) in 2012 (\citealt{harrison2013}) offers new opportunities to examine the hard X-ray properties of this source in detail. In early April 2016, \saxs\ went into outburst and exhibited the highest flux since \textit{NuSTAR}'s launch. 

This outburst provided an opportunity to search for cyclotron resonance scattering features (CRSFs), or cyclotron lines.  A CRSF was first discovered in Hercules X-1 by \cite{truemper1978}, and since then approximately 25 sources have been discovered with cyclotron absorption features ranging between 10 and 80 keV (\citealt{fuerst2016HEAD}). We discuss cyclotron lines in further detail in Section \ref{sec:absorption}.

In this work, we examine phase-averaged spectra of the two \textit{NuSTAR} observations shown in Figure \ref{fig:saxbatlc}, one taken during the precursor flare (hereafter Observation I) and one taken approximately 60 days later during outburst (hereafter Observation II). In Section \ref{sec:analysis}, we examine the spectral features of these two different flux states and notice changes in the pulse profile. We also perform pulse-phase spectroscopy on both observations in order to better characterize the accretion mechanisms driving the X-ray activity. In Section \ref{sec:absorption}, we discuss a potential absorption feature that could be related to cyclotron scattering.

\section{Observations and Data Reduction} \label{sec:observations}
\begin{figure*}
\includegraphics[scale=0.5]{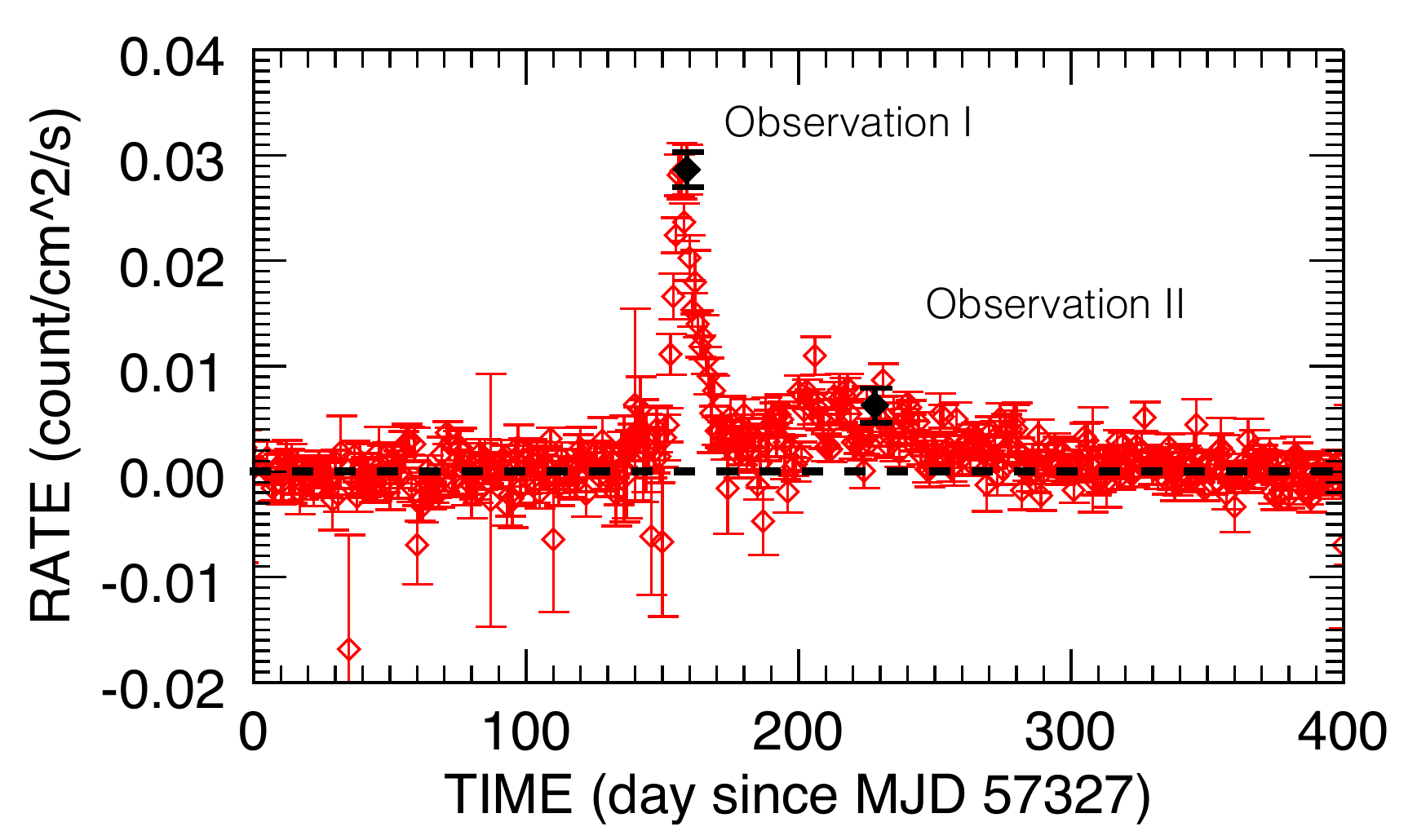}
\includegraphics[scale=0.5]{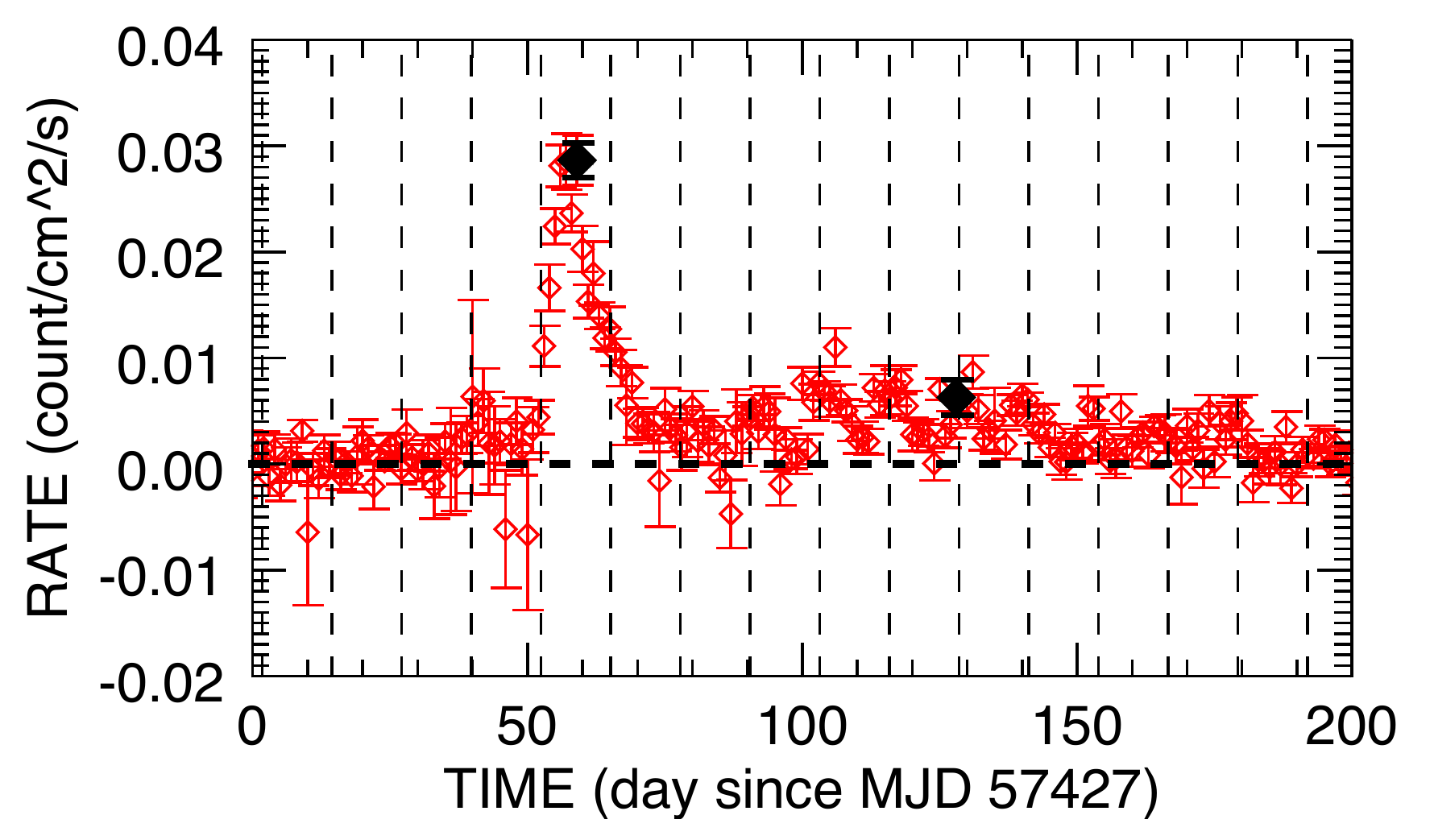}
\caption{Left: One day averaged light curve of \saxs\ taken with \textit{Swift}/BAT. The black points indicate the times of observation with \textit{NuSTAR}. The first observation on 8 April 2016 falls within the precursor flare, while the second took place on 16 June 2016, during the subsequent outburst. The black dashed line at rate=0 is plotted for reference. Right: Same as left for a narrower time range. Overplotted vertical dashed lines represent orbital phase 0 as defined by the \cite{baykal2000} ephemeris. While \cite{baykal2000} states that periastron passage occurs at $\phi_{\text{orbital}} \approx 0.4$, the alignment of the $\phi_{\text{orbital}}=0.0$ lines with the peaks in the \textit{Swift}/BAT light curve strongly suggests that periastron now coincides with orbital phase 0.}
\label{fig:saxbatlc}
\end{figure*}

The data for this work came from two Target of Opportunity (ToO) observations taken with \textit{NuSTAR}, which is sensitive to the 3--79 keV energy range (\citealt{harrison2013}). The time of observations are indicated in the left panel of Figure \ref{fig:saxbatlc}, which shows the one day averaged light curve of \saxs\ taken by \textit{Swift}/BAT. The first observation occurred on 8 April 2016 ($\phi_{\text{orbital}} \approx 0.6$) with an exposure time of 23 ks, and captured \saxs\ during its precursor flare. The second observation, with an exposure time of 41 ks, took place during the subsequent outburst period on 16 June 2016 when the source was at periastron passage ($\phi_{\text{orbital}} \approx 0.0$). The orbital phases were determined using the ephemeris found by \cite{baykal2000}. The right panel of Figure \ref{fig:saxbatlc} shows the light curve for this outburst overplotted with vertical lines representing orbital phase 0 as defined by \cite{baykal2000}. The alignment of orbital phase 0 with the peaks in the \textit{Swift}/BAT light curve suggest that periastron passage occurs at approximately orbital phase 0. This would imply a change to the orbital ephemeris since the work of \cite{baykal2000}, who found orbital phase 0.4 to corresponds to periastron passage of the neutron star.

We reduced these data using the {\fontfamily{qcr}\selectfont nupipeline} tool in NuSTARDAS 1.4.1 (HEAsoft version 6.16, CALDB 20161021). Source spectra were extracted from a circular region with an 120\arcsec\ radius centered on the source. Background spectra were extracted from a circular region with the same radius located on the other side of the field of view. The barycentric correction was applied using the {\fontfamily{qcr}\selectfont
barycorr} routine, and the photon arrival times were further corrected for the orbit of \saxs\ using the ephemeris from \cite{baykal2000}.

\section{Analysis and Results} \label{sec:analysis}

\subsection{Phase-averaged spectra} \label{sec:phaseavganalysis}

\begin{figure*}
\plottwo{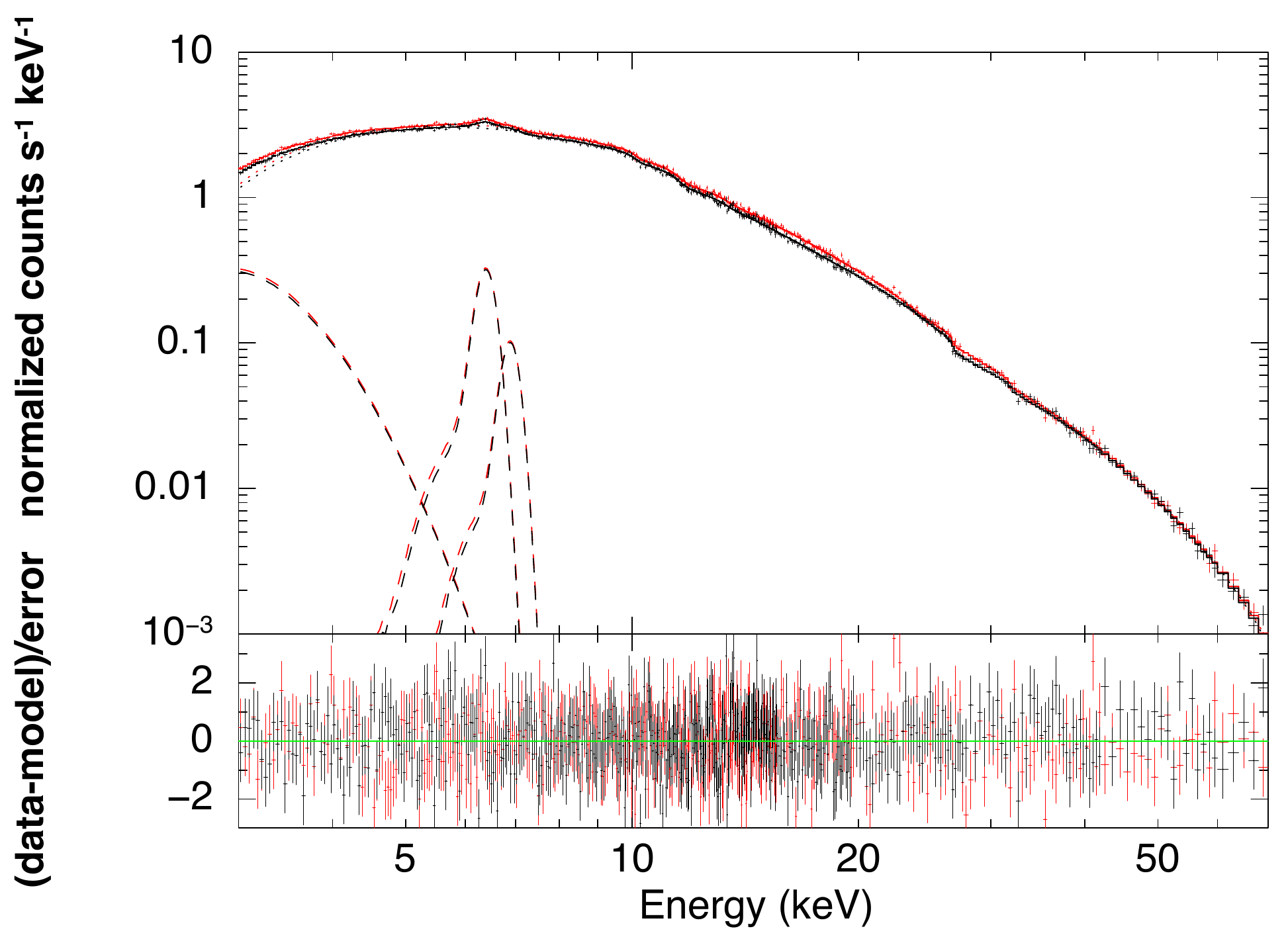} {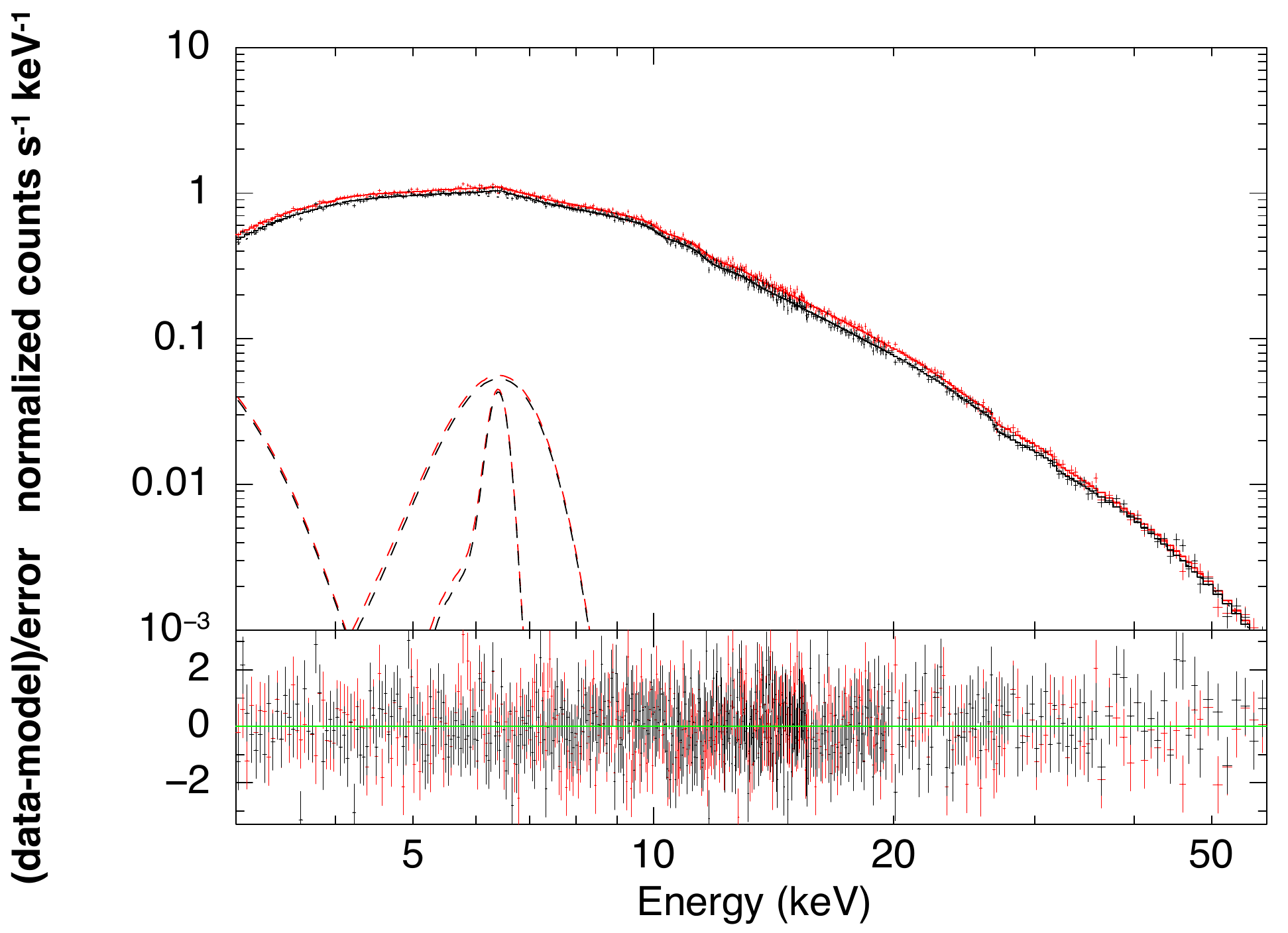}
\caption{The phase-averaged spectrum of \saxs\ for Observation I (left) and Observation II (right). Spectra from \textit{NuSTAR} FPMA are shown in red, and \textit{NuSTAR} FPMB in black. Both spectra are fitted with an NPEX model with a soft thermal component. In Observation I the best fit to the iron line features are two Gaussian emission lines, one at 6.4 and one at 6.9 keV. In Observation II the best fit is two Gaussians, one broad and one narrow, both centered at 6.4 keV. The bottom panel of each plot shows the data with the model subtracted, and divided by the error. The overall shape of the spectrum is consistent between the two observations; the only substantive difference is in the structure of the iron line.}
\label{fig:phaseavg}
\end{figure*}

Using the NuSTARDAS pipeline, we extracted a phase-averaged spectrum from the source region of each observation for both Focal Plane Modules A and B (FPMA and FPMB), which we rebinned using the FTOOLS command {\fontfamily{qcr}\selectfont grppha}. We rebinned the spectra above 10 keV using variable bin sizes to keep the errors in the residuals above 10 keV approximately constant. Several continuum models including a power law with a high energy cut-off (\citealt{white1983}), FDCut (Fermi-Dirac CUT-off power law, \citealt{tanaka1986}), and NPEX (Negative and Positive power law with an EXponential cut-off, e.g.\ \citealt{mihara1998}) were applied to the phase-averaged spectra using Xspec (version 12.8.2, \citealt{arnaud1996}). In Observation I, the best fit (\chisq\ of 857.10 for 915 degrees of freedom) was obtained for the NPEX model, which is defined in Xspec as
$$ f(E) = n_{1}(E^{-\alpha_{1}} + n_{2}E^{-\alpha_{2}}) e^{-E/kT},$$
however it should be noted that Xspec requires that $-10<\alpha_{2} < 0$, and the second power law has a positive exponent. The power law with a high energy cutoff failed to fit the hard energy tail of the data, and resulted in a higher \chisq\ of 881.17 for 918 degrees of freedom. Additionally, this model has been known to produce artificial absorption features around the cutoff energy which must be smoothed by the addition of Gaussians (e.g.\ \citealt{coburn2002}) or polynomials (e.g.\ \citealt{burderi2000}), thus adding additional free parameters to the model. The FDCut model also failed to fit the hard tail and resulted in a \chisq\ of 1217.53 for 918 degrees of freedom. Figure \ref{fig:phaseavg} shows the phase-averaged spectrum for each observation fitted with the best fit NPEX model.

In addition to the NPEX model, the data required a soft thermal component, two Gaussian emission lines, and a Gaussian absorption line that is discussed in Section \ref{sec:absorption}. Adding the blackbody component to the model reduced the \chisq\ from 910.23  to 857.10 for Observation I. In Observation II, the blackbody component is not as strongly significant, but we included it in the Obs.\ II model so that we can more directly compare the two observations.

In Observation I, narrow lines corresponding to the Fe K$\alpha$ line at 6.4 keV and a highly ionized iron line at 6.9 keV are present. In Observation II, the highly ionized iron line is no longer detected; instead, we find the best fit with a broad and a narrow line at 6.4 keV. We found that the narrow line is required by the data by running an f-test simulation using the tcl script {\fontfamily{qcr}\selectfont simftest} where the continuum model without the narrow Gaussian component was the null hypothesis. We created 500 simulated spectra and, using the test statistic \delchisq, found that the observed \delchisq\ was greater than the simulated values by an order of magnitude in all cases. We conclude that the narrow Fe line feature is strongly required by the data.

For both observations, the model was applied and fitted jointly to the \textit{NuSTAR} FPMA and FPMB spectra. The parameters in the model for the FPMB spectrum were tied to the parameters in the FPMA spectrum and related by a cross-calibration constant. Table \ref{table:npexparams} contains the results of the spectral fits shown in Figure \ref{fig:phaseavg}. In all models we used elemental abundances and cross sections given by \cite{wilms2000} and \cite{verner1996}, respectively. 

Using the X-ray and optical distance measurements for \sax, we calculated the 3 -- 40 keV luminosity of the Observation I to be between (0.5--1.1)$\times 10^{37}$ erg s$^{-1}$ and the 3 -- 40 keV luminosity of Observation II to be between (1.5--3.2)$\times 10^{36}$ erg s$^{-1}$, where the range in luminosities results from the different distance measurements to the source.

\begin{deluxetable}{lcc} [H]
\tablecolumns{3}
\tablecaption{Phase-averaged spectral parameters\tablenotemark{a}}
\tablewidth{0pt}
\tablehead{
\colhead{Parameter} & \colhead{Observation I} & \colhead{Observation II}  }  
\startdata
$N_{\text{H}}$ ($\times 10^{22}$ cm$^{-2}$) & 4.4 $\pm$ 1.4 &  2.9 $\pm$ 0.8 \\
$kT_{\text{BB}}$ (keV) & 0.35 $\pm$ 0.05 & 0.2 (fixed) \\
$n_{\text{BB}}$ (keV) & 0.008 $\pm$ 0.007 & 0.62 $\pm$ 0.04 \\
$\alpha_{1}$ & 0.61 $\pm$ 0.06 & 0.64 $\pm$ 0.1 \\
$\alpha_{2}$ & -2.1 $\pm$ 0.2 & -2.2 $\pm$ 0.3 \\
$n_{\alpha_{2}}$ & (8.2  $\pm$ 3.3)$\times 10^{-5}$ & (7.9 $\pm$ 7.0)$\times 10^{-5}$  \\ 
$kT_{\text{fold}}$ (keV) & 10.7 $\pm$ 0.8 & 9.1 $\pm$ 1.2 \\
log$_{10}$($F_{3-40 \text{ keV}}$) & -8.650 $\pm$ 0.004 & -9.191 $\pm$ 0.004 \\
E$_{\text{Fe}_{1}}$ (keV, fixed) & 6.4 & 6.4 \\
$\sigma_{\text{Fe}_{1}}$ (keV, fixed) & 0.1 & 0.05 \\
$n_{\text{Fe}_{1}}$  & (5.8 $\pm$ 0.5)$\times 10^{-4}$ & (6.9 $\pm$ 2.5)$\times 10^{-5}$ \\
$E_{\text{Fe}_{2}}$ (keV, fixed) & 6.9 & 6.4 \\
$\sigma_{\text{Fe}_{2}}$ (keV) & 0.1 (fixed) & 0.65 $\pm$ 0.15 \\
$n_{\text{Fe}_{2}}$  & (1.7 $\pm$ 0.5)$\times 10^{-4}$ & (3.2 $\pm$ 0.7)$\times 10^{-4}$ \\
$c_{\text{FPMA}}$ (fixed) & 1 & 1 \\
$c_{\text{FPMB}}$ & 1.021 $\pm$ 0.003 & 1.021 $\pm$ 0.004 \\
\chisq\ & 857.10 & 881.85 \\
Degrees of Freedom & 915 & 885
\enddata
\tablenotetext{a}{For the continuum model {\fontfamily{qcr}\selectfont constant * tbnew * (cflux * (npex * gabs)) + bbody + gauss + gauss)}. All errors are 90\% confidence intervals. The absorption feature is discussed in Section \ref{sec:absorption}.}
\label{table:npexparams}
\end{deluxetable}

\subsection{Pulse profiles} \label{sec:ppanalysis}

\begin{figure}
\centering
\includegraphics[trim={1cm 2cm 0.5cm 2.5cm},clip,scale=0.35]{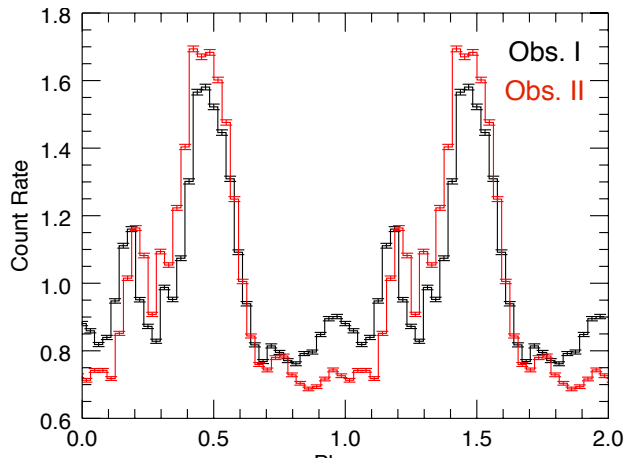}
\caption{Pulse profiles of \saxs\ for Observation I (black) and Observation II (red), shifted so that the peaks align. Between the two observations, the pulse profile changes from a three-peaked profile to a two-peaked profile.}
\label{fig:pulseprofs}
\end{figure}

The pulse period of \saxs\ was found using the FTOOL {\fontfamily{qcr}\selectfont efsearch} (\citealt{leahy1983}), which folds the light curve over a range of test periods and produces a chi-squared plot where the maximum is the best pulse period. The uncertainty in the pulse period was found by using the pulse profile to simulate 500 light curves, and finding the range of simulated pulse periods. We found a best pulse period of 349.23 $\pm$ 0.01\ s for Observation I and 349.20 $\pm$ 0.01\ s for Observation II. Pulse profiles were created using the FTOOL {\fontfamily{qcr}\selectfont efold}, which fold the light curve over the desired period to create a pulse profile. The pulse profiles are shown in Figure \ref{fig:pulseprofs}, where, for clarity, the phases have been chosen so that the peaks align.

\begin{figure*}
\plottwo{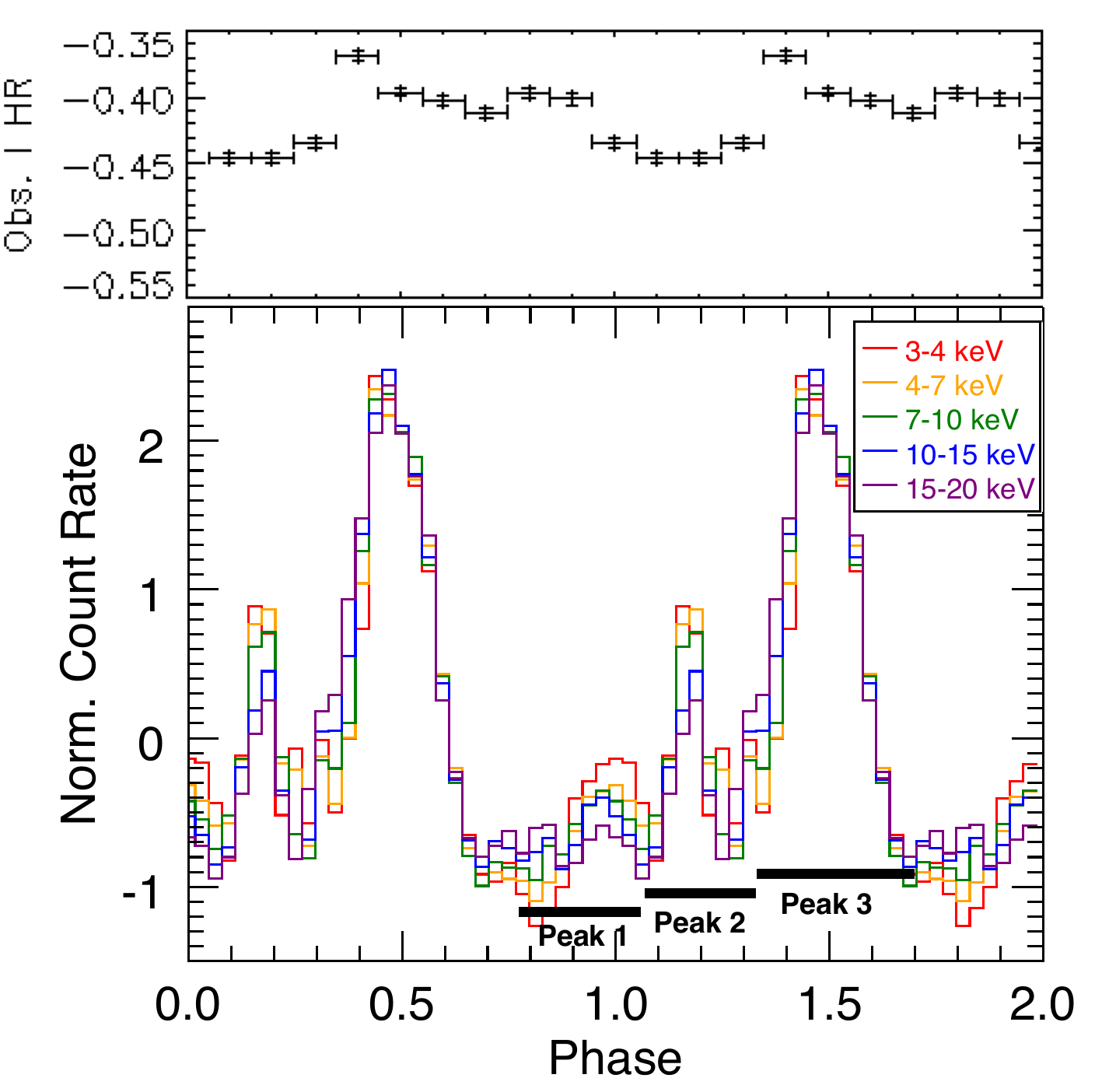}{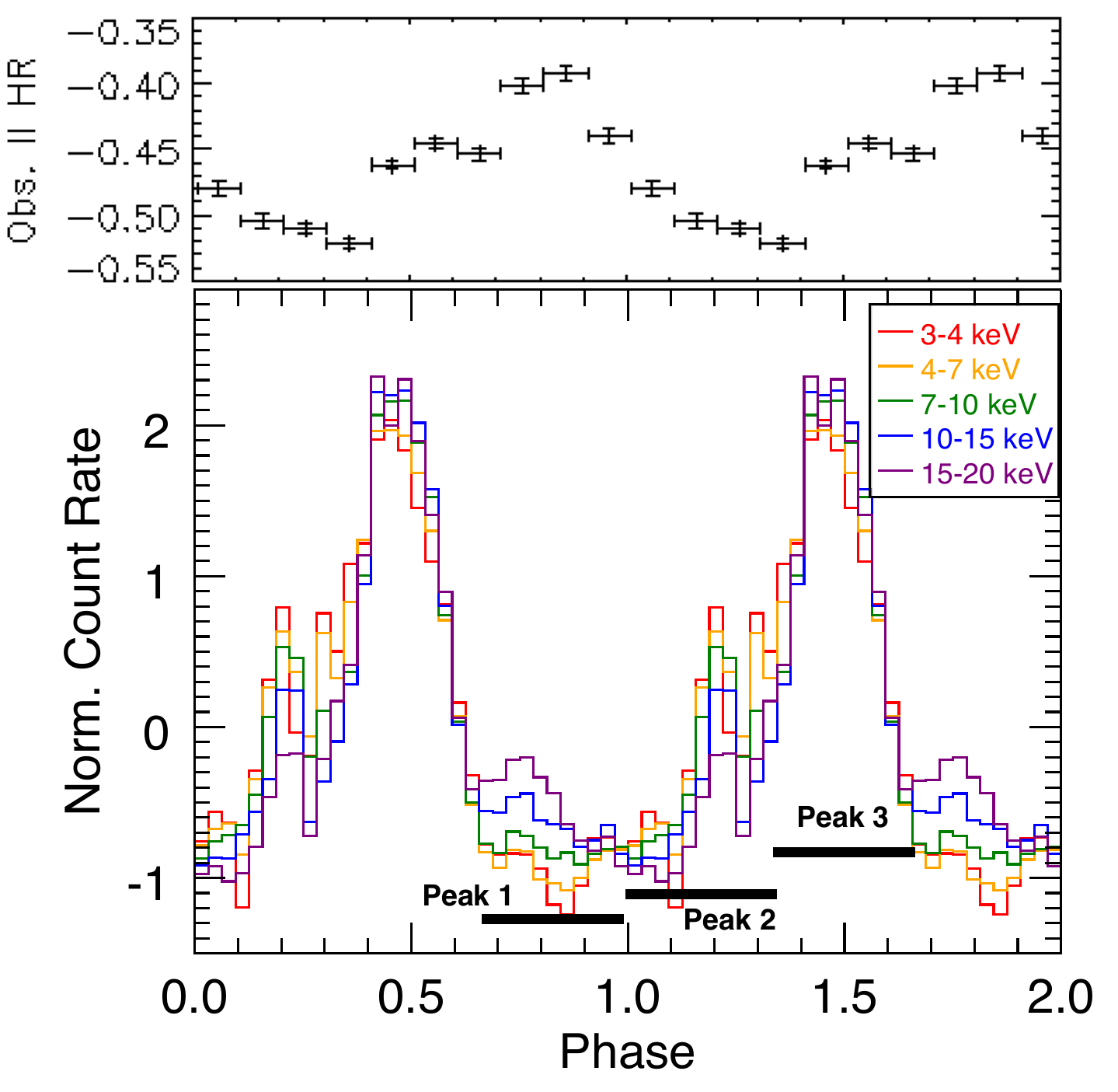}
\caption{Left: energy filtered pulse profiles for Observation I. Right: energy filtered pulse profiles for Observation II. The same energy ranges were used for both observations, and these bands were selected to demonstrate the changes in shape of the pulse profiles. The top panel of each figure shows the hardness ratios for 10 phase bins calculated as $(H-S)/(H+S)$ where $H$ is the count rate in the 10--20 keV band and $S$ is the count rate in the 3--10 keV band.}
\label{fig:enpulseprof}
\end{figure*}

The pulse profile for Observation I shows a three peaked profile where the peaks increase in count rate as a function of phase. In Observation II, the first peak is no longer visible, yielding a pulse profile with only two peaks. To check the energy dependence associated with these pulse profile changes, the cleaned event files for Observations I and II were filtered into five energy bands, and the pulse profiles were re-extracted in the same way as before. The energy filtered pulse profiles, shown in Figure \ref{fig:enpulseprof}, were normalized by subtracting the average count rate in each band and dividing by the respective standard deviation. In both observations, the strongest pulse peak is dominated by hard X-rays. In Observation I, the first peak is predominantly soft X-rays, and disappears entirely in the harder energy bands. The first peak in Observation II, which seemed to have disappeared in the full energy band pulse profile (Figure \ref{fig:pulseprofs}), is shown to still exist at harder energies, although it has also shifted in relative phase.

\subsection{Pulse-phase spectroscopy} \label{sec:phaseresanalysis}

The phase-averaged source event files were filtered into 10 equal phase bins with {\fontfamily{qcr}\selectfont xselect}. The phase-resloved spectra were extracted from these phase bins, and grouped with {\fontfamily{qcr}\selectfont grppha} to have 100 counts per spectral bin. We fitted each spectrum in the same way as the phase-averaged spectra using the same NPEX continuum model in the of 3 to 40 keV range. 

In initial fits to Observation I and Observation II, the \nh\ appeared to vary with pulse phase. These fluctuations were most likely caused by degeneracy in the model between the parameters for \nh\ and the values of the blackbody normalization. In order to limit this degeneracy, we fixed the \nh\ to its phase-averaged value in all phase bins.

\begin{figure*}
\centering
\includegraphics[scale=0.65]{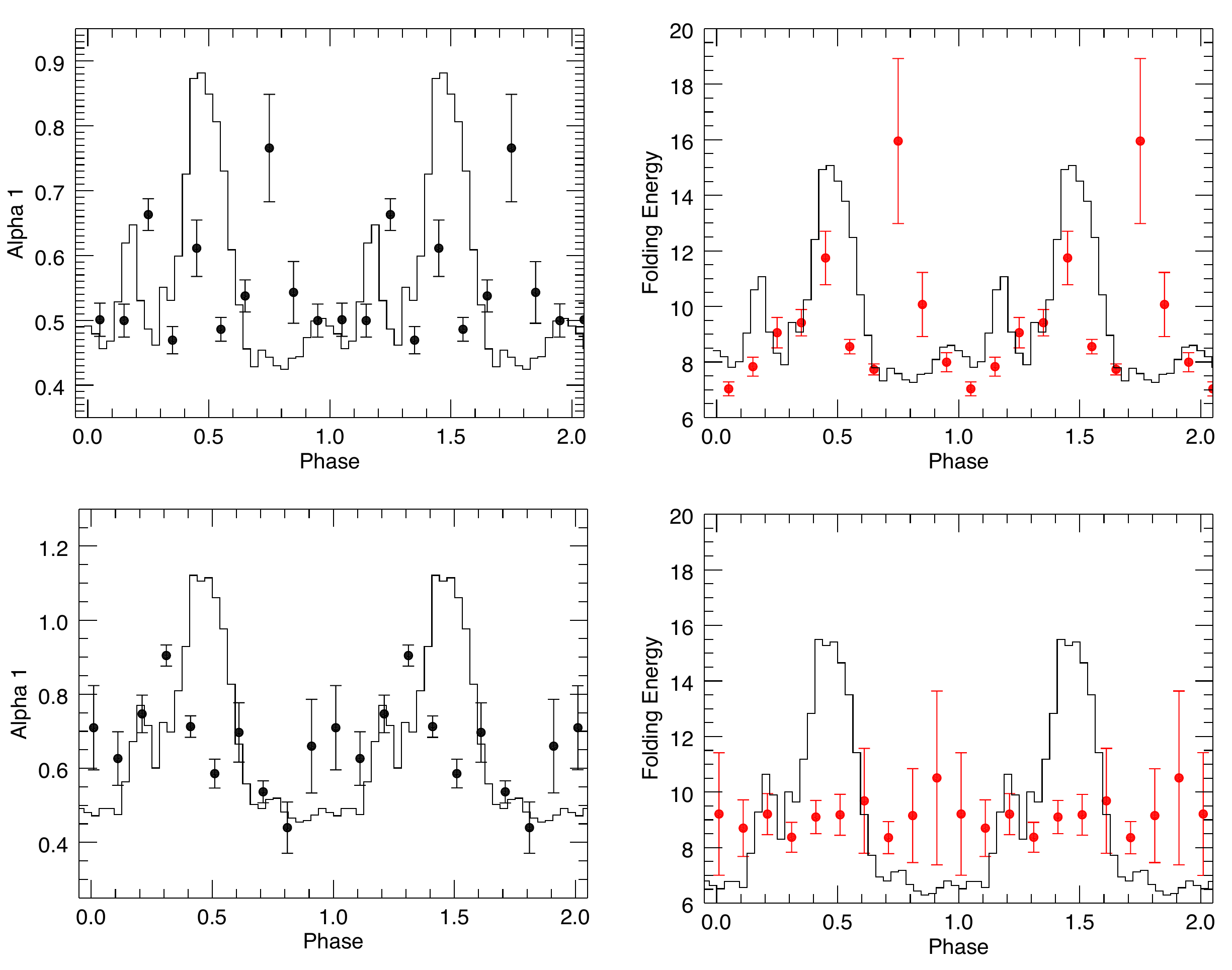}
\caption{The values of the NPEX parameters $\alpha_{1}$ and folding energy as a function of pulse phase for Observation I (top row) and Observation II (bottom row). The power law index $\alpha_{2}$ was fixed to the phase average value. The corresponding pulse profile, shown as a black line for reference in each panel, has been normalized to match the data. The change in folding energy from pulsed in Obs.\ I to consistent with constant in Obs.\ II suggests that there is still degeneracy between $\alpha_{1}$ and the folding energy. }
\label{fig:alphaphase}
\end{figure*}

Within each model the power law index $\alpha_{1}$, folding energy, flux, and normalizations of the blackbody component and Gaussian lines were variable. We fixed the power law index $\alpha_{2}$ to its phase-averaged value to reduce degeneracy between the two power law indices and the folding energy. Although it would have further reduced degeneracy in the model to fix the folding energy as well, we found that doing so resulted in a poor fit to the data. An error analysis of each variable parameter was also conducted in Xspec, yielding values with 90\% confidence intervals. To check for pulse phase dependence, these parameters were plotted against pulse phase (see Figure \ref{fig:alphaphase}).

In both observations, the line energies and widths for the two Gaussian components were fixed to their phase-averaged values because they could not be constrained in the phase-resolved analysis. In Observation I the iron line strengths are consistent with being constant with pulse phase. In Observation II, the iron lines are required by the data, as their absence results in reduced \chisq\ values of 1.2 or higher. However, due to the reduced signal to noise in these spectra and degeneracy in the continuum model, it is challenging to quantify the fluctuation of iron line strength with spin phase in a meaningful way.

The phase-resolved spectroscopy with 10 phase bins ultimately does not yield conclusive results about the behavior of spectral parameters with pulse phase. The lack of coherent pulsations in Figure \ref{fig:alphaphase} could be caused by degeneracies between the power law indices and the folding energy. 

\subsection{Pulse-peak spectroscopy} \label{sec:pulsepeakanalysis}

Since phase-resolved spectroscopy using fine phase bins did not yield conclusive results, we used the shape of the pulse profile to motivate a different binning selection. Both observations were divided into three phase bins, each of which covered a pulse peak within the pulse profile (see black lines in Figure \ref{fig:enpulseprof}). These phase bins were labeled Peak 1, Peak 2, and Peak 3, respectively, where Peak 3 is the strongest pulse in the pulse profile. We extracted three spectra for each observation according to this phase selection, and use the same binning and energy range restrictions as in the phase-averaged analysis.

For both observations, we fit the spectra from the three pulse-peak phase bins simultaneously and refer to parameters that are tied together, so as to be the same value in all phase bins, as global parameters (as in \cite{ballhausen2017}, see method description in \cite{kuhnel2015,kuhnel2016}). We also fixed the Fe line energies and widths to their phase-averaged values in all phase bins to reduce the number of free parameters. We fitted each of the three pulse peak spectra with the continuum model used on the phase-averaged spectra ({\fontfamily{qcr}\selectfont constant * tbnew (bbody + cflux * npex + gaus + gaus)}). We defined the column density \nh\, the folding energy, and the FPMB normalization to be global parameters since we expect these to vary little with pulse phase. Additionally, the blackbody temperature is a global parameter in Observation I, however this parameter is poorly constrained in Observation II and had to be fixed to its phase-averaged value in all phase bins.

We found the following global parameters for Observations I and II, respectively: the \nh\ values are $(4.6 \pm 1.4) \times 10^{22} \text{ cm}^{-2} $ and $(3.0 \pm 0.7)\times 10^{22} \text{ cm}^{-2}$, the folding energies are $10.6 \pm 0.6$ keV and $8.0 \pm 0.5$ keV, and the FPMB normalizations are $1.021 \pm 0.003$ and $1.021 \pm 0.004$. The blackbody temperature for Observation I is $0.37 \pm 0.04$ keV. The continuum parameters, found in Table \ref{table:pulsepeakparams}, generally do not vary strongly with pulse phase.

\begin{figure*}
\centering
\includegraphics[scale=0.65]{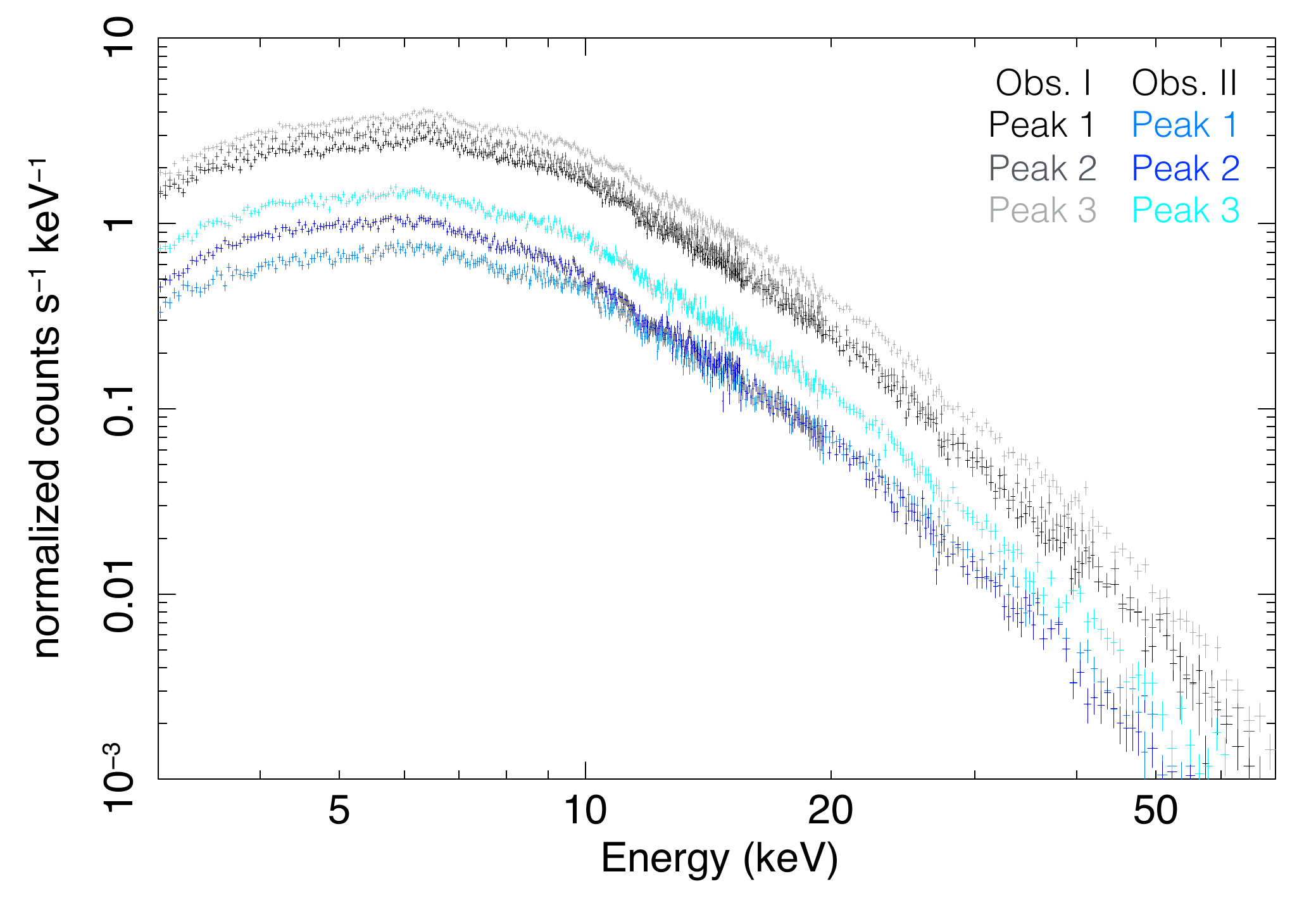}
\caption{Spectra extracted from the three pulse-peak phase bins noted in Figure \ref{fig:enpulseprof} for Observation I (greyscale) and Observation II (blues). Only the FPMA spectrum from each pulse peak is plotted, for clarity. While there are some differences in overall count rate, the slope of the power law is generally consistent across all three phase bins.}
\label{fig:peakspectra}
\end{figure*}

In Observation I, Figure \ref{fig:enpulseprof} indicates that there are only slight variations in continuum shape between pulse peaks. For Observation I, the pulse profiles indicate that Peak 1 and Peak 2 are both softer than Peak 3. However, as seen in Figure \ref{fig:peakspectra}, the steepness of the power law does not change significantly across the three pulse peaks. The continuum parameters for Obs.\ I (Table \ref{table:pulsepeakparams}) are generally consistent with being constant.

In Observation II, the largest change with pulse phase can be seen in the $\alpha_{1}$ parameter, which varies significantly between Peak 1 and Peak 2. This change can be seen in Figure \ref{fig:peakspectra}, where the Obs.\ II Peak 1 and Obs.\ II Peak 2 spectra have different slopes. The change in slope could also be driven by the relative strength of the second power law in NPEX, which is significantly stronger in Obs.\ II Peak 1 than in the other two phase bins.

\begin{deluxetable*}{llccc}
\tablecolumns{5}
\tablecaption{Pulse-peak-resolved spectral parameters\tablenotemark{a,b}}  
\tablewidth{0pt}
\centering
\tablehead{
\colhead{Observation} & \colhead{Parameter} & \colhead{Peak 1} & \colhead{Peak 2} & \colhead{Peak 3}  }  
\startdata
Observation I & $n_{\text{BB}}$ & 0.007 $\pm$ 0.004 & 0.004 $\pm$ 0.003 & 0.010 $\pm$ 0.005 \\
\hspace{0.2cm} & $\alpha_{1}$  & 0.63 $\pm$ 0.06 & 0.70 $\pm$ 0.06 & 0.55 $\pm$ 0.06  \\
\hspace{0.2cm} & $\alpha_{2}$  &  -2.1 $\pm$ 0.2 & -2.0 $\pm$ 0.2  & -2.2 $\pm$ 0.2 \\
\hspace{0.2cm} & $n_{\alpha_{2}}$ & (6.1 $\pm\ 4.9)\times 10^{-5}$ & (9.0 $\substack{+7.7 \\ -4.2})\times 10^{-5} $ & (6.9 $\pm\ 3.4)\times 10^{-5}$ \\
\hspace{0.2cm} & log$_{10}$($F_{3-40 \text{ keV}}$) & -8.755 $\pm$ 0.005 & -8.689 $\pm$ 0.005 & -8.562 $\pm$ 0.004 \\
\hspace{0.2cm} & $n_{\text{Fe}_{1}}$ &(6.1 $\pm\ 0.8)\times 10^{-4} $ & (5.5 $\pm\ 1.0)\times 10^{-4} $ & (5.8 $\pm\ 0.8)\times 10^{-4} $\\
\hspace{0.2cm} & $n_{\text{Fe}_{2}}$ &(1.7 $\pm\ 0.7)\times 10^{-4}$ & (2.0 $\pm\ 1.0)\times 10^{-4}$ & (1.7 $\pm\ 0.8)\times 10^{-4}$ \\
Observation II & $n_{\text{BB}}$ & 0.12 $\pm$ 0.03 & 0.03 $\pm$ 0.03 & 0.10 $\pm$ 0.05 \\
\hspace{0.2cm} & $\alpha_{1}$  & 0.45 $\pm$ 0.07  & 0.65 $\pm$ 0.07 & 0.54 $\pm$ 0.07 \\
\hspace{0.2cm} & $\alpha_{2}$ & -2.5 $\pm$ 0.2  & -2.7 $\pm$ 0.2  &  -2.5 $\pm$ 0.2 \\
\hspace{0.2cm} & $n_{\alpha_{2}}$ & (1.1 $\substack{+0.6 \\ -0.3})\times 10^{-4} $  & (2.1 $\substack{+1.3 \\ -0.8})\times 10^{-6} $ & (5.7 $\substack{+3.0 \\ -1.9})\times 10^{-5} $\\
\hspace{0.2cm} & log$_{10}$($F_{3-40 \text{ keV}}$) & -9.315 $\pm$ 0.004  &  -9.279 $\pm$ 0.004 & -9.056 $\pm$ 0.003 \\
\hspace{0.2cm} & $n_{\text{Fe}_{\text{broad}}}$ & (1.6 $\pm\ 0.8)\times 10^{-4}$ & (4.3 $\pm\ 1.0)\times 10^{-4}$ & (3.2 $\pm\ 1.2)\times 10^{-4}$\\
\hspace{0.2cm} & $n_{\text{Fe}_{\text{narrow}}}$ & (7.5 $\pm\ 3.4)\times 10^{-5}$ & (6.8 $\pm\ 4.1)\times 10^{-5}$ & (6.1 $\pm\ 5.2)\times 10^{-5}$
\enddata
\tablenotetext{a}{All errors are 90\% confidence intervals. The energy and width of the Fe lines were fixed to their phase-averaged values in all phase bins. The \nh, $kT_{\text{BB}}$, folding energy, and FPMB normalization were defined as global parameters.}
\tablenotetext{b}{The reduced \chisq\ values were 1.029 and 1.005 for Observations I and II, respectively.} 
\label{table:pulsepeakparams}
\end{deluxetable*}

\section{A possible absorption feature} \label{sec:absorption}

\subsection{Introduction}

\textit{NuSTAR}'s hard X-ray sensitivity also allows us to search the hard X-ray spectrum for cyclotron lines. The CRSF appears as an absorption-like feature at hard energies that occurs due to resonant scattering of continuum photons by electrons on quantized Landau energy levels in the strong magnetic field of the neutron star's accretion column. Because the energy spacing between Landau levels can be given by $E_{\text{CRSF}}=\hbar e B / m_{e}c = 11.57 B_{12}$ keV (where $B_{12}$ is the magnetic field strength in units of $10^{12}$\,G), the line energy at which the CRSF appears is a direct measurement of the strength of the pulsar's magnetic field close to the neutron star surface. Understanding the magnetic properties of neutron star X-ray binaries like \saxs\ is necessary for modeling both accretion and emission models (e.g.\ \citealt{becker2007}).

\subsection{Phase-averaged absorption feature}

Despite the lack of an obvious CRSF in the phase-averaged spectra, we attempted to find an upper limit on a cyclotron line feature at any energy. We added the Xspec Gaussian absorption model component {\fontfamily{qcr}\selectfont gabs} to the continuum model and stepped the line energy from 8 to 60 keV. In each observation, we found an improvement in \chisq\ at specific energies ($\sim$30 keV for Obs.\ I and $\sim$15 keV for Obs.\ II).

We verified this \delchisq\ by running {\fontfamily{qcr}\selectfont steppar} on the absorption line energy while the line strength and continuum parameters (NPEX: $\alpha$1, $\alpha$2, folding energy, cross normalization; cflux: log of flux) were variable. All other parameters in the model were fixed to their phase-averaged values. Since {\fontfamily{qcr}\selectfont steppar} returns the \chisq\ for the model fit as it steps through $E_{\text{CRSF}}$, we compared the maximum and minimum \chisq\ to obtain a second measurement of \delchisq. 

\begin{figure*}
\plotone{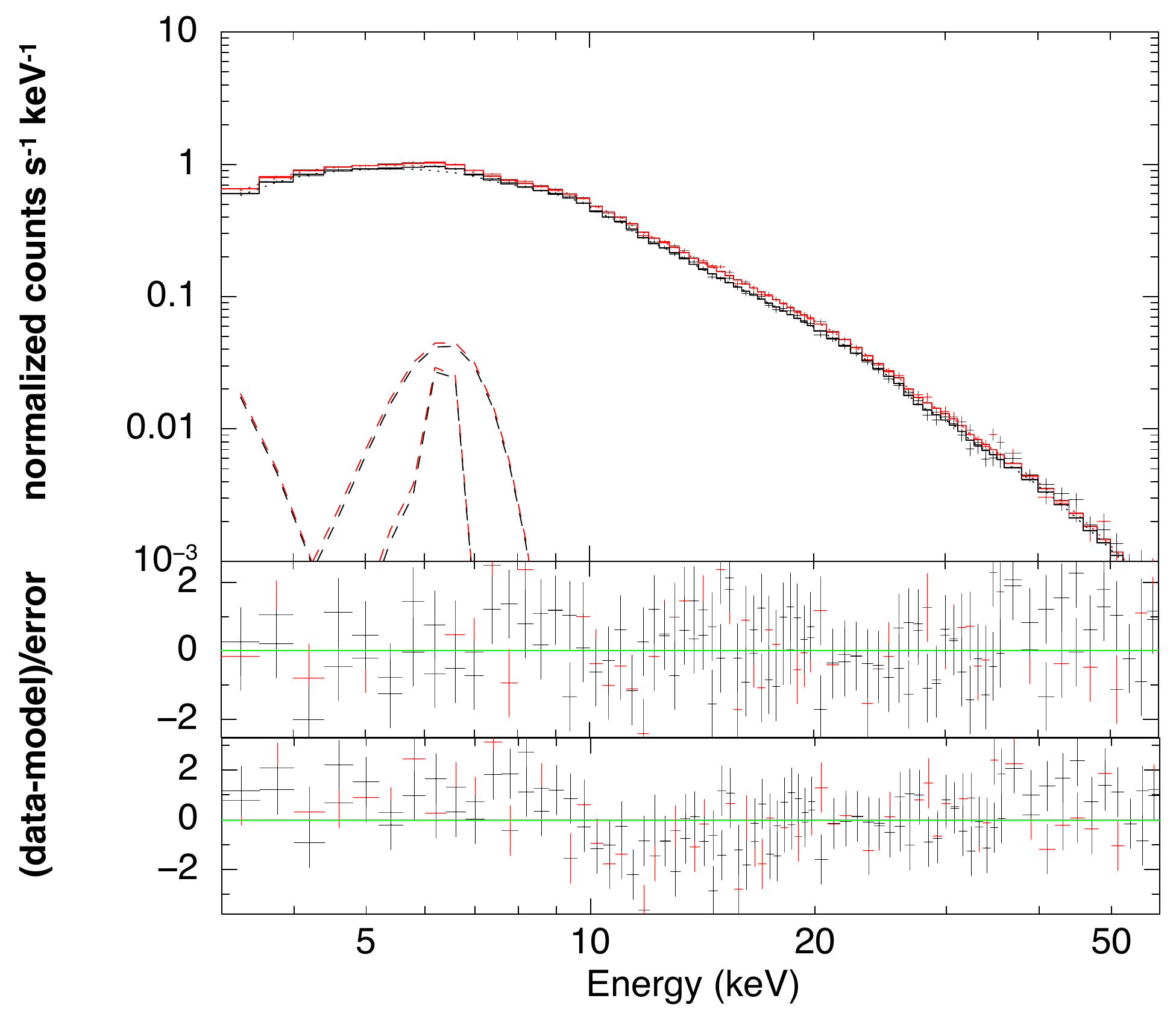}
\caption{Top: the \nustar\ FPMA (red) and FPMB (black) spectrum from the Observation II Peak 2 phase bin. The spectrum has been fitted with the best fit model, which includes {\fontfamily{qcr}\selectfont gabs}, and binned more strongly to demonstrate the fit. Middle: the data with the model subtracted and divided by the error for the best fit model shown above. Bottom: the data with the model subtracted and divided by the error for the best fit model in which the {\fontfamily{qcr}\selectfont gabs} component has been removed. The dip in {\fontfamily{qcr}\selectfont delchi} at approximately 12 keV is the absorption feature.}
\label{fig:peak2withwithoutgabs}
\end{figure*}

In the Obs. I phase-averaged spectrum, the best fit absorption feature had a line energy of 31 keV and a depth of 0.2. The line width was frozen to 4 keV, a typical width for CRSFs (e.g.\ \citealt{tendulkar2014}), due to this parameter being poorly constrained. This line was detected with a \delchisq\ of 2.5, which is not strong enough to be considered significant. In the Obs. II phase-averaged spectrum, the best fit absorption feature had a line energy of 15.3 keV, a depth of 0.3, and a width of 3.3 keV, and resulted in a \delchisq\ of 6.4.

We used the Xspec tcl script {\fontfamily{qcr}\selectfont simftest} to determine the significance of the absorption feature in the Obs. II spectrum. We used the continuum NPEX model without {\fontfamily{qcr}\selectfont gabs} as our null hypothesis, and simulated fake spectra from this model. {\fontfamily{qcr}\selectfont Simftest} then fitted the continuum NPEX model with and without {\fontfamily{qcr}\selectfont gabs} and recorded the change in \chisq\ for 10,000 trials. As a test statistic, we used a ratio of the normalized \chisq\ for the model fit without the absorption feature to the normalized \chisq\ for the model fit with the absorption feature, where the normalized \chisq\ is the \chisq\ divided by the associated degrees of freedom (in other words, the random variate of the $F$-distribution. We found that the observed normalized \chisq\ ratio was greater than 96.2\%\ of the ratios from the sample of 10,000 simulated data sets, corresponding to $\sim1.8\sigma$ (one-tail test).

\subsection{Pulse-peak absorption feature}
The 10 phase-resolved spectra had too few counts in each bin to effectively search for a weak absorption feature. We therefore checked for phase dependence in the feature using the three pulse-peak spectra from each observation. For both observations, as was done with the phase-averaged spectra, we stepped the energy of the absorption feature from 8 to 60 keV and performed a fit at each energy. For both observations, we found that the absorption lines in the phase bins Peak 1 and Peak 3 were not significant detections. However, the absorption features in Peak 2 were significant in both observations, and the line parameters were similar. In Observation 1, the best fit Peak 2 absorption feature had a central energy of 11.2 keV, a line width of 4 keV, and a depth of 0.5. In Observation II, the best fit Peak 2 feature had a central energy of 12 keV, a width of 3.3 keV, and a depth of 0.7. In both of these cases, the widths were taken from the phase-averaged spectral line fits. The line parameters for each phase bin and the \delchisq\ improvement are listed in Table \ref{table:pulsepeakcrsf}. We found that adding an absorption feature resulted in a \chisq\ improvement of 11 in Observation I and 34 in Observation II. In Figure \ref{fig:peak2withwithoutgabs}, we show the fit to the Obs.\ II Peak 2 spectrum with and without the {\fontfamily{qcr}\selectfont gabs} model component. The absorption feature can be seen as a slight dip in the residuals in the bottom panel.

We ran {\fontfamily{qcr}\selectfont simftest} and calculated the normalized \chisq\ ratio in the same way as for the phase-averaged spectrum to test the significance of the absorption features in the Peak 2 phase bin of both observations. We found that, for Observation I, our observed normalized \chisq\ ratio was greater than 99.2\%\ of the ratios from the simulated spectra. In Observation II, the observed ratio was greater than 99.8\%\ of ratios from the simulated spectra. 

The $\approx2.4\sigma$ and $\approx2.9\sigma$ (one-tail test) presence of this absorption feature within the same phase bin of both observations, and at approximately the same energy, suggests that this feature is not a calibration error or statistical phenomenon.

\begin{deluxetable*}{llccc}
\tablecolumns{5}
\tablecaption{Pulse-peak-resolved spectral parameters\tablenotemark{a}}  
\tablewidth{0pt}
\centering
\tablehead{
\colhead{Observation} & \colhead{Parameter} & \colhead{Peak 1} & \colhead{Peak 2} & \colhead{Peak 3}  }  
\startdata
Observation I & $E_{{\text{\fontfamily{qcr}\selectfont gabs}}}$ (keV) &  12.4 & 11.4  & 29   \\
\hspace{0.2cm} & $d_{{\text{\fontfamily{qcr}\selectfont gabs}}}$  & 0.2 $\substack{+0.3 \\ -0.2}$  & 1.0 $\pm$ 0.4  & 0.2 $\substack{+0.3 \\ -0.2}$  \\
\hspace{0.2cm} & $w_{{\text{\fontfamily{qcr}\selectfont gabs}}}$ (keV, fixed)  &  4 & 4  & 4 \\
\hspace{0.2cm} & Maximum $\Delta \chi^{2}$ from adding {\fontfamily{qcr}\selectfont gabs} & 0.4  & 11  & 0.8 \\
\hspace{0.2cm} & Significance & \nodata & 99.2\% & \nodata \\
Observation II & $E_{\text{{\fontfamily{qcr}\selectfont gabs}}}$ (keV) &  \nodata & 12.0 $\pm$ 0.9  & 16.1 $\substack{+1.3 \\ -1.5}$ \\
\hspace{0.2cm} & $d_{{\text{\fontfamily{qcr}\selectfont gabs}}}$ & \nodata  & 0.8 $\pm$ 0.3  &  0.4 $\pm$ 0.2 \\
\hspace{0.2cm} & $w_{{\text{\fontfamily{qcr}\selectfont gabs}}}$ (keV, fixed) & \nodata  & 3.3  & 3.3  \\
\hspace{0.2cm} & Maximum $\Delta \chi^{2}$ from adding {\fontfamily{qcr}\selectfont gabs} & \nodata  & 34  & 6 \\
\hspace{0.2cm} & Significance & \nodata &  99.8\% & \nodata 
\enddata
\tablenotetext{a}{Where no uncertainty is given, these values were found to be the best fit in Xspec, but needed to be frozen to ensure a stable error analysis.}
\label{table:pulsepeakcrsf}
\end{deluxetable*}

\section{Discussion} \label{sec:discussion}

Through our phase-averaged spectral analysis we found that the power-law continuum of \saxs\ did not significantly change between the two observations, despite their difference in flux. Visual inspection of the spectra do not reveal obvious signs of a cyclotron line feature, however the addition of a Gaussian absorption feature does result in an improvement of \chisq\ in both observations. We found this change in \chisq\ is only statistically significant in Obs.\ II, where it resulted a probability of chance improvement of the \chisq\ of 3.8\%. In both observations, this absorption feature shows phase dependence and is only significantly detected in the phase bin labeled Peak 2 in Figure \ref{fig:enpulseprof} (in a one-tail test, the significance corresponds to $\approx 2.4\sigma$ and $\approx 2.9\sigma$ for Observations I and II, respectively). The strong phase dependence of this feature suggests that it could be a CRSF, which have been known to vary in strength and energy with pulse phase (e.g.\ \citealt{staubert2013,schwarm2017b}). 

There remains a small possibility that this feature is a calibration issue with \textit{NuSTAR}, which has previously known calibration uncertainties in the range of 10-14 keV (e.g. \citealt{fuerst2013}). However, more recent updates to the \textit{NuSTAR} database have significantly improved the calibration (\citealt{madsen2015}). Additionally, the phase dependences of the absorption feature suggests that it is not a calibration issue. There is also an unexplained feature in the spectra of some neutron star X-ray binaires known as the 10 keV bump (see Section 6.4 in \cite{coburn2002} for a review). This feature can be present either in emission or absorption at $\approx$10 keV and can appear in sources with or without cyclotron features. For pulsars with low energy CRSFs, emission in this region is generally caused by Comptonized cyclotron cooling (e.g.\ \citealt{ferrigno2009}), however it is unclear what, if any, physical mechanism drives this feature (\citealt{coburn2002}).

If the absorption feature found in our data is caused by cyclotron scattering, we can constrain the magnetic field of \saxs. The magnetic field strength can be found using the equation $E_{\text{CRSF}}=11.57 B_{12} (1+z)^{-1}$\ keV, where $B_{12}$ is the magnetic field strength in units of $10^{12}$ G. Using the best fit value of $12.0 \pm 0.9$ keV found in the Observation II Peak 2 phase bin, we estimate a magnetic field of $\approx 1 \times 10^{12} (1+z)^{-1}$ G. This would categorize \saxs, along with KS 1947$+$300, which has $E_{\text{CRSF}}=12.5$ keV (\citealt{fuerst2014}), and 4U 0115$+$63, with $E_{\text{CRSF}}=11-15$ keV (\citealt{iyer2015}), as a weakly magnetic compared to the majority of cyclotron sources (\citealt{fuerst2013}). 

In both Observation I, a soft thermal component is necessary to describe the continuum below 5 keV (\delchisq $\approx$ 50). We include the blackbody component in Observation II \cite{hickox2004} describes the origin of the soft X-ray excess in pulsars, and for moderate luminosity pulsars such as \saxs, attributes this excess to either disk reprocessing, diffuse gas emission, or a combination of both. The presence of ionized iron indicates that disk reprocessing is the most likely cause of the soft excess in this source. 

We detect a narrow Fe K$\alpha$ feature at 6.4 keV in both of our observations. Previous works have detected the Fe K$\alpha$ line in \saxs\ both during outburst with RXTE (e.g.\ \citealt{baykal2002}) and quiescence with \textit{Chandra} and \textit{XMM-Newton} (e.g.\ \citealt{reig2014,gimenez2015}). Each of these works finds a faint, narrow ($\sigma \sim$ 0.1 keV) line at 6.4 keV. Our data provides the clearest spectral picture of the iron line complex to date.

During the bright precursor flare, we detect a highly ionized iron line at 6.9 keV. This is the first time this line has been observed in \saxs. The structure of this iron line appears to change between the two observations. In Observation II, the 6.9 keV line feature is no longer visible, and the spectrum is best fit by a broad and narrow Gaussian component, each centered at 6.4 keV. This change could possibly be driven by the differences in the accretion flow between the two observations. The precursor flare is believed to be a significant accretion event (\citealt{camero2014}) where temperatures in the disk could result in highly ionized states of iron. After the precursor flare, more moderate accretion occurs during periastron passage, which could explain the lack of a highly ionized iron line in Observation II. The 0.65\ keV width of the broad line that appears in Obs.\ II is consistent with orbital velocities of gas at the edge of the magnetosphere.

Although the period of pulsation is constant over Observation I and Observation II, the shape of the pulse profile changes from three peaks to two peaks. Figure \ref{fig:enpulseprof} indicates that the changes in shape are energy dependent. These differences could also be a result of differences in the accretion rate between the two observations. In order to fully describe the shape of the pulse profile, light bending from the opposite pulsar beam must also be considered (e.g.\ \citealt{falkner2016}). 

Phase-resolved spectroscopy with 10 equal phase bins proved inconclusive, partially because of reduced signal to noise in these narrow phase bins. The broader pulse-peak-resolved spectroscopy revealed that, while the strongest pulse peak was the hardest, the overall continuum shape did not change significantly. Some fluctuations in the power law indices are visible, but degeneracy between these parameters make the variations difficult to interpret. 

\section{Summary}

We have observed \saxs\ for the first time with \textit{NuSTAR} during its precursor flare and outburst. While the flux of Observation I is greater than that of Observation II, the shape of the continuum does not change significantly between the two observations. The phase averaged spectra are best fit by the NPEX model where $\alpha_{1}$ is $\sim\! 0.66$ and $\alpha_{2}$ is $\sim -2.0$ for both observations. The best fit also includes a soft blackbody component and several ionized iron lines. 

These observations provide the most detailed picture of the iron line complex in \saxs\ to date. In addition to detecting the Fe K$\alpha$ line in both observations, we also observe a highly ionized iron line during the bright precursor flare. The iron line structure appears to change from two narrow gaussians at 6.4 and 6.9 keV in Observation I to a narrow and broad component, both at 6.4 keV, in Observation II. While our spectral resolution is not sufficient to determine the precise nature of these changes, differences in the accretion rate between the precursor flare and regular outburst could result in different temperatures of the accretion disk, and thus the ionization level of iron.

Pulse-phase spectroscopy with 10 equal phase bins does not lead to conclusive results about the spectral variations in this source. The iron line normalizations show only weak fluctuations with pulse phase, and the $\alpha$ parameters driving the slope of the power law do not show smooth variations with pulse phase. Pulse-peak resolved spectroscopy with three phase bins dividing up the peaks of the pulse profile indicate that the overall shape of the continuum does not vary strongly with pulse phase. 

We possibly detect an absorption feature at $\sim$ 12 keV in both observations. This feature exhibits strong phase dependence, and could be a CRSF. Future observations with better spectral resolution in the 12 keV range could help prove the presence of this feature. If this feature is found to be a CRSF, it would classify \saxs, along with KS 1947$+$300 and 4U 0115$+$63, as one of the lowest magnetic field cyclotron line sources.

\acknowledgements
We would like to thank the anonymous referee for his or her helpful contributions to the final paper. This work was supported by NASA grant number NNX15AV32G. We would like to thank the NuSTAR Galactic Binaries Science Team for comments and contributions. This research made use of NuSTARDAS, developed by ASDC (Italy) and Caltech (USA).

\bibliography{my_bib}

\begin{thebibliography}{}
\expandafter\ifx\csname natexlab\endcsname\relax\def\natexlab#1{#1}\fi

\bibitem[{{Arnaud}(1996)}]{arnaud1996}
{Arnaud}, K.~A. 1996, in Astronomical Society of the Pacific Conference Series,
  Vol. 101, Astronomical Data Analysis Software and Systems V, ed. G.~H.
  {Jacoby} \& J.~{Barnes}, 17

\bibitem[{{Ballhausen} {et~al.}(2017){Ballhausen}, {Pottschmidt}, {F{\"u}rst},
  {Wilms}, {Tomsick}, {Schwarm}, {Stern}, {Kretschmar}, {Caballero},
  {Harrison}, {Boggs}, {Christensen}, {Craig}, {Hailey}, \&
  {Zhang}}]{ballhausen2017}
{Ballhausen}, R., {Pottschmidt}, K., {F{\"u}rst}, F., {et~al.} 2017, ArXiv
  e-prints, arXiv:1707.05648

\bibitem[{{Baykal} {et~al.}(2007){Baykal}, {Inam}, {Stark}, {Heffner},
  {Erkoca}, \& {Swank}}]{baykal2007}
{Baykal}, A., {Inam}, S.~{\c C}., {Stark}, M.~J., {et~al.} 2007, \mnras, 374,
  1108

\bibitem[{{Baykal} {et~al.}(2000){Baykal}, {Stark}, \& {Swank}}]{baykal2000}
{Baykal}, A., {Stark}, M.~J., \& {Swank}, J. 2000, \apjl, 544, L129

\bibitem[{{Baykal} {et~al.}(2002){Baykal}, {Stark}, \& {Swank}}]{baykal2002}
{Baykal}, A., {Stark}, M.~J., \& {Swank}, J.~H. 2002, \apj, 569, 903

\bibitem[{{Becker} \& {Wolff}(2007)}]{becker2007}
{Becker}, P.~A., \& {Wolff}, M.~T. 2007, \apj, 654, 435

\bibitem[{{Burderi} {et~al.}(2000){Burderi}, {Di Salvo}, {Robba}, {La Barbera},
  \& {Guainazzi}}]{burderi2000}
{Burderi}, L., {Di Salvo}, T., {Robba}, N.~R., {La Barbera}, A., \&
  {Guainazzi}, M. 2000, \apj, 530, 429

\bibitem[{{Camero} {et~al.}(2014){Camero}, {Zurita}, {Guti{\'e}rrez-Soto},
  {{\"O}zbey Arabac{\i}}, {Nespoli}, {Kiaeerad}, {Beklen},
  {Garc{\'{\i}}a-Rojas}, \& {Caballero-Garc{\'{\i}}a}}]{camero2014}
{Camero}, A., {Zurita}, C., {Guti{\'e}rrez-Soto}, J., {et~al.} 2014, \aap, 568,
  A115

\bibitem[{{Coburn} {et~al.}(2002){Coburn}, {Heindl}, {Rothschild}, {Gruber},
  {Kreykenbohm}, {Wilms}, {Kretschmar}, \& {Staubert}}]{coburn2002}
{Coburn}, W., {Heindl}, W.~A., {Rothschild}, R.~E., {et~al.} 2002, \apj, 580,
  394

\bibitem[{{Ducci} {et~al.}(2014){Ducci}, {Jourdain}, {Wilms}, \&
  {Bozzo}}]{ducci2014}
{Ducci}, L., {Jourdain}, E., {Wilms}, J., \& {Bozzo}, E. 2014, The Astronomer's
  Telegram, 6154

\bibitem[{{Ducci} {et~al.}(2008){Ducci}, {Sidoli}, {Paizis}, {Mereghetti}, \&
  {Pizzochero}}]{ducci2008}
{Ducci}, L., {Sidoli}, L., {Paizis}, A., {Mereghetti}, S., \& {Pizzochero},
  P.~M. 2008, in Proceedings of the 7th INTEGRAL Workshop, 116

\bibitem[{{Falkner} {et~al.}(2016){Falkner}, {Schwarm}, {Wolff}, {Becker}, \&
  {Wilms}}]{falkner2016}
{Falkner}, S., {Schwarm}, F.-W., {Wolff}, M.~T., {Becker}, P.~A., \& {Wilms},
  J. 2016, in AAS/High Energy Astrophysics Division, Vol.~15, AAS/High Energy
  Astrophysics Division, 201.08

\bibitem[{{Ferrigno} {et~al.}(2009){Ferrigno}, {Becker}, {Segreto}, {Mineo}, \&
  {Santangelo}}]{ferrigno2009}
{Ferrigno}, C., {Becker}, P.~A., {Segreto}, A., {Mineo}, T., \& {Santangelo},
  A. 2009, \aap, 498, 825

\bibitem[{{Filippova} {et~al.}(2004){Filippova}, {Lutovinov}, {Shtykovsky},
  {Revnivtsev}, {Burenin}, {Arefiev}, {Pavlinsky}, \&
  {Sunyaev}}]{filippova2004}
{Filippova}, E.~V., {Lutovinov}, A.~A., {Shtykovsky}, P.~E., {et~al.} 2004,
  Astronomy Letters, 30, 824

\bibitem[{{Fuerst} \& {MAGNET Collaboration}(2016)}]{fuerst2016HEAD}
{Fuerst}, F., \& {MAGNET Collaboration}. 2016, in AAS/High Energy Astrophysics
  Division, Vol.~15, AAS/High Energy Astrophysics Division, 201.07

\bibitem[{{F{\"u}rst} {et~al.}(2013){F{\"u}rst}, {Grefenstette}, {Staubert},
  {Tomsick}, {Bachetti}, {Barret}, {Bellm}, {Boggs}, {Chenevez}, {Christensen},
  {Craig}, {Hailey}, {Harrison}, {Klochkov}, {Madsen}, {Pottschmidt}, {Stern},
  {Walton}, {Wilms}, \& {Zhang}}]{fuerst2013}
{F{\"u}rst}, F., {Grefenstette}, B.~W., {Staubert}, R., {et~al.} 2013, \apj,
  779, 69

\bibitem[{{F{\"u}rst} {et~al.}(2014){F{\"u}rst}, {Pottschmidt}, {Wilms},
  {Kennea}, {Bachetti}, {Bellm}, {Boggs}, {Chakrabarty}, {Christensen},
  {Craig}, {Hailey}, {Harrison}, {Stern}, {Tomsick}, {Walton}, \&
  {Zhang}}]{fuerst2014}
{F{\"u}rst}, F., {Pottschmidt}, K., {Wilms}, J., {et~al.} 2014, \apjl, 784, L40

\bibitem[{{Gim{\'e}nez-Garc{\'{\i}}a}
  {et~al.}(2015){Gim{\'e}nez-Garc{\'{\i}}a}, {Torrej{\'o}n}, {Eikmann},
  {Mart{\'{\i}}nez-N{\'u}{\~n}ez}, {Oskinova}, {Rodes-Roca}, \&
  {Bernab{\'e}u}}]{gimenez2015}
{Gim{\'e}nez-Garc{\'{\i}}a}, A., {Torrej{\'o}n}, J.~M., {Eikmann}, W., {et~al.}
  2015, \aap, 576, A108

\bibitem[{{Harrison} {et~al.}(2013){Harrison}, {Craig}, {Christensen},
  {Hailey}, {Zhang}, {Boggs}, {Stern}, {Cook}, {Forster}, {Giommi},
  {Grefenstette}, {Kim}, {Kitaguchi}, {Koglin}, {Madsen}, {Mao}, {Miyasaka},
  {Mori}, {Perri}, {Pivovaroff}, {Puccetti}, {Rana}, {Westergaard}, {Willis},
  {Zoglauer}, {An}, {Bachetti}, {Barri{\`e}re}, {Bellm}, {Bhalerao},
  {Brejnholt}, {Fuerst}, {Liebe}, {Markwardt}, {Nynka}, {Vogel}, {Walton},
  {Wik}, {Alexander}, {Cominsky}, {Hornschemeier}, {Hornstrup}, {Kaspi},
  {Madejski}, {Matt}, {Molendi}, {Smith}, {Tomsick}, {Ajello}, {Ballantyne},
  {Balokovi{\'c}}, {Barret}, {Bauer}, {Blandford}, {Brandt}, {Brenneman},
  {Chiang}, {Chakrabarty}, {Chenevez}, {Comastri}, {Dufour}, {Elvis}, {Fabian},
  {Farrah}, {Fryer}, {Gotthelf}, {Grindlay}, {Helfand}, {Krivonos}, {Meier},
  {Miller}, {Natalucci}, {Ogle}, {Ofek}, {Ptak}, {Reynolds}, {Rigby},
  {Tagliaferri}, {Thorsett}, {Treister}, \& {Urry}}]{harrison2013}
{Harrison}, F.~A., {Craig}, W.~W., {Christensen}, F.~E., {et~al.} 2013, \apj,
  770, 103

\bibitem[{{Hickox} {et~al.}(2004){Hickox}, {Narayan}, \&
  {Kallman}}]{hickox2004}
{Hickox}, R.~C., {Narayan}, R., \& {Kallman}, T.~R. 2004, \apj, 614, 881

\bibitem[{{Hulleman} {et~al.}(1998){Hulleman}, {in 't Zand}, \&
  {Heise}}]{hullemann1998}
{Hulleman}, F., {in 't Zand}, J.~J.~M., \& {Heise}, J. 1998, \aap, 337, L25

\bibitem[{{Iyer} {et~al.}(2015){Iyer}, {Mukherjee}, {Dewangan}, {Bhattacharya},
  \& {Seetha}}]{iyer2015}
{Iyer}, N., {Mukherjee}, D., {Dewangan}, G.~C., {Bhattacharya}, D., \&
  {Seetha}, S. 2015, \mnras, 454, 741

\bibitem[{{K{\"u}hnel} {et~al.}(2015){K{\"u}hnel}, {M{\"u}ller}, {Kreykenbohm},
  {Schwarm}, {Grossberger}, {Dauser}, {Pottschmidt}, {Ferrigno}, {Rothschild},
  {Klochkov}, {Staubert}, \& {Wilms}}]{kuhnel2015}
{K{\"u}hnel}, M., {M{\"u}ller}, S., {Kreykenbohm}, I., {et~al.} 2015, Acta
  Polytechnica, 55, 123

\bibitem[{{K{\"u}hnel} {et~al.}(2016){K{\"u}hnel}, {Falkner}, {Grossberger},
  {Ballhausen}, {Dauser}, {Schwarm}, {Kreykenbohm}, {Nowak}, {Pottschmidt},
  {Ferrigno}, {Rothschild}, {Mart{\'{\i}}nez-N{\'u}{\~n}ez}, {Torrej{\'o}n},
  {F{\"u}rst}, {Klochkov}, {Staubert}, {Kretschmar}, \& {Wilms}}]{kuhnel2016}
{K{\"u}hnel}, M., {Falkner}, S., {Grossberger}, C., {et~al.} 2016, Acta
  Polytechnica, 56, 41

\bibitem[{{Leahy} {et~al.}(1983){Leahy}, {Elsner}, \& {Weisskopf}}]{leahy1983}
{Leahy}, D.~A., {Elsner}, R.~F., \& {Weisskopf}, M.~C. 1983, \apj, 272, 256

\bibitem[{{Lutovinov} {et~al.}(2003){Lutovinov}, {Molkov}, \&
  {Revnivtsev}}]{lutovinov2003}
{Lutovinov}, A.~A., {Molkov}, S.~V., \& {Revnivtsev}, M.~G. 2003, Astronomy
  Letters, 29, 713

\bibitem[{{Madsen} {et~al.}(2015){Madsen}, {Harrison}, {Markwardt}, {An},
  {Grefenstette}, {Bachetti}, {Miyasaka}, {Kitaguchi}, {Bhalerao}, {Boggs},
  {Christensen}, {Craig}, {Forster}, {Fuerst}, {Hailey}, {Perri}, {Puccetti},
  {Rana}, {Stern}, {Walton}, {J{\o}rgen Westergaard}, \& {Zhang}}]{madsen2015}
{Madsen}, K.~K., {Harrison}, F.~A., {Markwardt}, C.~B., {et~al.} 2015, \apjs,
  220, 8

\bibitem[{{Manousakis} {et~al.}(2007){Manousakis}, {Reig}, \&
  {Kougentakis}}]{manousakis2007}
{Manousakis}, A., {Reig}, P., \& {Kougentakis}, A. 2007, The Astronomer's
  Telegram, 1085

\bibitem[{{Mihara} {et~al.}(1998){Mihara}, {Makishima}, \&
  {Nagase}}]{mihara1998}
{Mihara}, T., {Makishima}, K., \& {Nagase}, F. 1998, Advances in Space
  Research, 22, 987

\bibitem[{{Reig}(2004)}]{reig2004}
{Reig}, P. 2004, in ESA Special Publication, Vol. 552, 5th INTEGRAL Workshop on
  the INTEGRAL Universe, ed. V.~{Schoenfelder}, G.~{Lichti}, \& C.~{Winkler},
  373

\bibitem[{{Reig}(2011)}]{reig2011}
{Reig}, P. 2011, \apss, 332, 1

\bibitem[{{Reig} {et~al.}(2014){Reig}, {Doroshenko}, \& {Zezas}}]{reig2014}
{Reig}, P., {Doroshenko}, V., \& {Zezas}, A. 2014, \mnras, 445, 1314

\bibitem[{{Reig} {et~al.}(2010){Reig}, {S{\l}owikowska}, {Zezas}, \&
  {Blay}}]{reig2010}
{Reig}, P., {S{\l}owikowska}, A., {Zezas}, A., \& {Blay}, P. 2010, \mnras, 401,
  55

\bibitem[{{Rivinius} {et~al.}(2013){Rivinius}, {Carciofi}, \&
  {Martayan}}]{rivinius2013}
{Rivinius}, T., {Carciofi}, A.~C., \& {Martayan}, C. 2013, \aapr, 21, 69

\bibitem[{{Schwarm} {et~al.}(2017){Schwarm}, {Ballhausen}, {Falkner},
  {Sch{\"o}nherr}, {Pottschmidt}, {Wolff}, {Becker}, {F{\"u}rst},
  {Marcu-Cheatham}, {Hemphill}, {Sokolova-Lapa}, {Dauser}, {Klochkov},
  {Ferrigno}, \& {Wilms}}]{schwarm2017b}
{Schwarm}, F.-W., {Ballhausen}, R., {Falkner}, S., {et~al.} 2017, \aap, 601,
  A99

\bibitem[{{Sguera} {et~al.}(2012){Sguera}, {Drave}, {Goossens}, {Bird},
  {Sidoli}, {Fiocchi}, {Bazzano}, \& {Tarana}}]{sguera2012}
{Sguera}, V., {Drave}, S., {Goossens}, M., {et~al.} 2012, The Astronomer's
  Telegram, 4168

\bibitem[{{Sidoli} {et~al.}(2005){Sidoli}, {Mereghetti}, {Larsson},
  {Chernyakova}, {Kreykenbohm}, {Kretschmar}, {Paizis}, {Santangelo},
  {Ferrigno}, \& {Falanga}}]{sidoli2005}
{Sidoli}, L., {Mereghetti}, S., {Larsson}, S., {et~al.} 2005, \aap, 440, 1033

\bibitem[{{Staubert} {et~al.}(2013){Staubert}, {Klochkov}, {Vasco}, {Postnov},
  {Shakura}, {Wilms}, \& {Rothschild}}]{staubert2013}
{Staubert}, R., {Klochkov}, D., {Vasco}, D., {et~al.} 2013, \aap, 550, A110

\bibitem[{Tanaka(1986)}]{tanaka1986}
Tanaka, Y. 1986, in Radiation Hydrodynamics in Stars and Compact Objects
  (Berlin, Heidelberg: Springer Berlin Heidelberg), 198--221

\bibitem[{{Tendulkar} {et~al.}(2014){Tendulkar}, {F{\"u}rst}, {Pottschmidt},
  {Bachetti}, {Bhalerao}, {Boggs}, {Christensen}, {Craig}, {Hailey},
  {Harrison}, {Stern}, {Tomsick}, {Walton}, \& {Zhang}}]{tendulkar2014}
{Tendulkar}, S.~P., {F{\"u}rst}, F., {Pottschmidt}, K., {et~al.} 2014, \apj,
  795, 154

\bibitem[{{Truemper} {et~al.}(1978){Truemper}, {Pietsch}, {Reppin}, {Voges},
  {Staubert}, \& {Kendziorra}}]{truemper1978}
{Truemper}, J., {Pietsch}, W., {Reppin}, C., {et~al.} 1978, \apjl, 219, L105

\bibitem[{{Verner} {et~al.}(1996){Verner}, {Ferland}, {Korista}, \&
  {Yakovlev}}]{verner1996}
{Verner}, D.~A., {Ferland}, G.~J., {Korista}, K.~T., \& {Yakovlev}, D.~G. 1996,
  \apj, 465, 487

\bibitem[{{White} {et~al.}(1983){White}, {Swank}, \& {Holt}}]{white1983}
{White}, N.~E., {Swank}, J.~H., \& {Holt}, S.~S. 1983, \apj, 270, 711

\bibitem[{{Wilms} {et~al.}(2000){Wilms}, {Allen}, \& {McCray}}]{wilms2000}
{Wilms}, J., {Allen}, A., \& {McCray}, R. 2000, \apj, 542, 914

\bibitem[{{Wolff} {et~al.}(2016){Wolff}, {Becker}, {Gottlieb}, {F{\"u}rst},
  {Hemphill}, {Marcu-Cheatham}, {Pottschmidt}, {Schwarm}, {Wilms}, \&
  {Wood}}]{wolff2016}
{Wolff}, M.~T., {Becker}, P.~A., {Gottlieb}, A.~M., {et~al.} 2016, \apj, 831,
  194

\end{thebibliography}

\end{document}